\documentclass[final,5p,times,twocolumn]{elsarticle}

\usepackage{amssymb}

\usepackage{lineno}

\usepackage{braket}
\usepackage{bm}
\usepackage{amsmath}
\usepackage{color}
\usepackage{ulem}

\newcommand{\nucl}[2]{{}^{#1}\mathrm{#2}}
\newcommand{\CC}{\nucl{12}{C} + \nucl{12}{C}}

\newcommand{\ane}{\alpha + \nucl{20}{Ne}}
\newcommand{\pna}{p + \nucl{23}{Na}}
\newcommand{\er}{E_\mathrm{R}}

\journal{Physics Letters B}

\begin{document}

\begin{frontmatter}



\title{Impact of the molecular resonances on the $\CC$ fusion reaction rate}


\author[kosen,riken]{Yasutaka Taniguchi}
\author[riken]{Masaaki Kimura}

\affiliation[kosen]{organization={Department of Information Engineering, National Institute of Technology (KOSEN), Kagawa College},
            addressline={551 Kohda, Takuma-cho}, 
            city={Mitoyo},
            postcode={769-1192}, 
            state={Kagawa},
            country={Japan}}

\affiliation[riken]{organization={RIKEN Nishina Center},
            addressline={2-1 Hirosawa}, 
            city={Wako},
            postcode={351-0198}, 
            state={Saitama},
            country={Japan}}

\begin{abstract}
 The properties of the low-energy $\CC$ molecular resonances, which potentially enhance the fusion
 reaction rate at low temperatures, have been investigated by a full-microscopic nuclear model
 employing various  nuclear energy density functionals. We show that some 
 density functionals plausibly describe the observed high-spin $\CC$ molecular resonances and predict many
 $0^+$ and $2^+$ resonances at low energies, which enhance the reaction rate. We also discuss how
 the uncertainty in the nuclear energy density functionals propagates to that of the reaction rate.
\end{abstract}




\end{frontmatter}




\section{Introduction}
The $\CC$ fusion reaction plays a vital role in explosive phenomena in the universe, such as X-ray
superbursts (XRSBs)~\cite{Cumming2001,Strohmayer2002}, type Ia supernovae~\cite{Arnett1969,Mori2019}, and the evolution of massive
stars~\cite{RevModPhys.74.1015}.  Its reaction rate at low temperatures has been of great interest for many
years. Recently, direct measurements have reached to the energies as low as $E\simeq
2$~MeV~\cite{Spillane2007,Fruet2020,Tan2020}. In addition, an indirect measurement by the Trojan Horse Method (THM)~\cite{Tumino2018}
has reported that resonances below 2~MeV might significantly enhance the
reaction rate. However, the reaction rate at low temperatures still
remains inconclusive due to the uncertainties in the measurements.

Consequently, the reaction rates extrapolated to low temperatures are used in the simulations
of astrophysical phenomena. The standard estimation by Caughlan and Fowler (CF88)~\cite{Caughlan1988} assumes
a constant modified astrophysical spectroscopic factor ($S^\ast$-factor), whereas the hindrance
model~\cite{Jiang2007} asserted its suppression  at low temperatures. Contrary, Cooper et al.~\cite{Cooper2009}
discussed an increased reaction rate at low temperatures.
Thus, there are ad-hoc variations of the reaction rate, and hence, it is crucial to
impose a constraint by the nuclear model calculations. In particular, it is an essential task
for nuclear theories to investigate whether the resonances increase the reaction rate.

In our previous work~\cite{Taniguchi2021}, on the basis of  the antisymmetrized molecular dynamics (AMD) calculations, we
discussed the low-energy resonances which affect the $\CC$ fusion 
reaction. Here, we conduct extensive research, and report the low-energy resonances which have the
$\CC$ molecule-like structure and significantly affect the reaction rage. Employing several nuclear
density functionals, we assess the constraint on the reaction rate imposed by the nuclear theory. 

\section{Theoretical Model}
The description of the low-energy resonances requires an accurate description of various reaction
channels and rearrangement of nucleons. To meet this requirement, we employ the generator coordinate
method (GCM)~\cite{Hill1953,Griffin1957} which describes a resonance by a superposition of the wave functions for
various channels,  
\begin{align}
 \Psi^{J} = \sum_{ciK}f_{ciK}P^{J}_{MK}\Psi_c(d_i)
 + \sum_{iK}g_{iK}P^{J}_{MK}\Psi_{\rm Mg}(\beta_i),
\end{align}
where $P^{J}_{MK}$ is the angular momentum projector, and $\Psi_c(d_i)$ is the channel wave
function where the subscript $c$ denotes the channels relevant to the fusion reaction, i.e., $\CC$, $\ane$, and $\pna$ channels. 
The inter-nuclear distance $d_i$ ranges from 0.5 to 8~fm for the $\CC$ , from 0.5 to 9~fm for the $\ane$, and from 0.5 to 7~fm for the $\pna$, with the common intervals of 0.5~fm.
In addition, the wave functions of the compound nucleus $^{24}{\rm Mg}$, denoted by  $\Psi_{\rm Mg}(\beta_i)$, are
also superposed to describe the coupling between the entrance and exit channels and the compound
states. The resonance energy and the coefficients $f_{ciK}$ and $g_{iK}$ are determined by the
diagonalization of the Hamiltonian defined by the density functionals.
We have employed several
parameter sets of the density functionals, the Gogny parameter sets (D1S~\cite{Berger1991} and
D1M*~\cite{gonzalez2018}) and Skyrme parameter sets (SkM*~\cite{Bartel1982}, SLy4~\cite{Chabanat1998,Chabanat1998e}, and SIII~\cite{Beiner1975}). 

Each channel wave function is the Slater determinant of the nucleon wave packets projected to
positive parity, 
\begin{equation}
  \Psi  = \frac{1+ P_r}{2}{\mathcal{A}}\{\varphi_1\cdots\varphi_A\},\\
\end{equation} 
where $P_r$ is the parity operator, and each nucleon is described by Gaussian wave packet~\cite{Kanada-Enyo2012},
\begin{equation}
  \varphi_p = \prod_{\sigma=x,y,z}
e^{-\nu_\sigma\left(r_\sigma - Z_{p\sigma}\right)^2}
 \left(a_p\ket{\uparrow}+b_p\ket{\downarrow}\right)\eta_p,
\end{equation}
where isospin $\eta_p$ is fixed to either proton or neutron for all nucleons. The parameter
$\bm{Z}_p$ represents the mean position and momentum of each nucleon.  By controlling the parameter
$\bm{Z}_p$, the rearrangement of nucleons can be described in an unified manner. In the
practical calculations of the  $\CC$, $\alpha+\nucl{20}{Ne}$, and $\pna$ channels,  $\bm{Z}_p$, spin 
($a_p$ and $b_p$), and width parameters $\bm{\nu}$ are determined by the energy variation with the
constraint on the inter-nuclear distance $d_i$~\cite{Taniguchi2004}. Through the variational calculations, the
rotation and polarization of nuclei are naturally described depending on the inter-nuclear distance.
The wave functions for the compound nucleus $^{24}{\rm Mg}$  are calculated by the energy variation
with the constraint on the quadrupole deformation parameter $\beta$~\cite{PhysRevC.56.1844}.

\begin{figure}[tbp]
\begin{center}
  \includegraphics[width=0.475\textwidth]{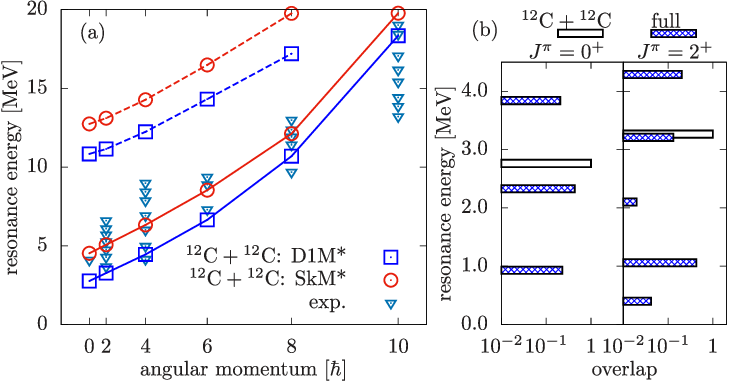}
\end{center} 
\caption{
 (a) The energies of the $\CC$ molecular resonances calculated by
 using the D1M* and SkM* functionals with only the $\CC$ channel . The lines connect the states with
 strong E2  transitions. The observed data~\cite{Almqvist1960,Almqvist1963,Mazarakis1973,Cosman1975,Basrak1976,Erb1976,Fletcher1976,Wada1977,Fortune1977,Ohkubo1982} are also shown for comparison. 
 (b) The energies of the $\CC$ molecular resonances calculated by using the D1M* functional
 without (white bars) and with (blue bars) the channel coupling. The lengths of the blue bars
 indicate the overlap of the wave functions with the $\CC$ channel.
 }\label{molres}
\end{figure}

\section{Results and Discussions}
 Figure~\ref{molres}(a) shows the $\CC$ molecular bands calculated
with only the $\CC$ channel, neglecting the coupling with the exit channels and compound states.  For comparison, we show the results obtained by using the 
D1M* and SkM* functionals. Both density functionals yield two molecular bands, and the lower
one qualitatively reproduces the observed resonances~\cite{Almqvist1960,Almqvist1963,Mazarakis1973,Cosman1975,Basrak1976,Erb1976,Fletcher1976,Wada1977,Fortune1977,Ohkubo1982} from low to high spin. The D1M*
yields lower resonance energies than the SkM*, and this trend is common to other parameter sets,
i.e., the Gogny functionals always yield lower resonance energies than the Skyrme functionals.  
This may be due to the finite-range two-body interaction of the Gogny functionals, which gives
stronger attraction between two nuclei at large distances. As discussed later, this makes the
qualitative difference in the low-temperature reaction rate.

Now, we discuss the channel coupling effects. Figure~\ref{molres}(b) shows how the $\CC$ molecular
states are fragmented into many resonances by the channel coupling in the case of the D1M*
functional.  The white bars show the energies of the $J^\pi=0^+$ and $2^+$ resonances obtained with
only the $\CC$ channel, whereas the blue  bars show the resonances obtained by including all
channels. The lengths of blue bars indicate the magnitude of the overlap between the wave functions obtained by including all channels and those obtained by including only the $\CC$ channel (white bars). It is clear that the $\CC$ molecular states are
fragmented into many resonances due to the channel coupling. Note that some of the
resonances are located down to 1~MeV within the Gamow window of the XRSBs. In contrast, Skyrme
functionals do not yield the resonances  as low as 1~MeV, because the energies of the $\CC$
molecular states without the channel coupling is too high.  

We apply the $R$-matrix theory~\cite{Lane1958} and the Breit-Wigner formula~\cite{Breit1936} to evaluate
resonance parameters and $S^\ast$-factors. We calculate the reduced width amplitude
(RWA)~\cite{kanada2014,Chiba2017} which is the overlap between the decay channel and resonance wave
functions,    
\begin{align}
 &a^2y_l^{A_1 + A_2}(a)& \nonumber\\
 &= \sqrt{
  \left(
   \begin{array}{c}
    24\\
    A_1\\
   \end{array}
   \right)
  }
 \Braket{\delta(r-a)[\Phi_{A_1}\Phi_{A_2}Y_{l}(\hat r)]^J_{M}|
 \Psi^{J\pi}_M}, \label{eq:rwa1}
\end{align}
where the  wave function for the decay channel is composed of nuclei with masses $A_1$ and $A_2$
separated by 
the distance $a$ with the orbital angular momentum $l$.  We consider six decay channels,
$\alpha+\nucl{20}{Ne}(0^+_1)$, $\alpha+\nucl{20}{Ne^*}(2^+_1)$, $\ane^\ast(4_1^+)$, $p+\nucl{23}{Na}
(3/2^+_1)$, $p+\nucl{23}{Na^*}(5/2^+_1)$, and $\pna^\ast(7/2_1^+)$, which  are denoted by
$\alpha_0$, $\alpha_1$, $\alpha_2$, $p_0$, $p_1$, and $p_2$, respectively. We also calculated
$\alpha_3$ and $\alpha_4$ channels and found that they are negligible. The wave functions
of $\alpha$, $\nucl{20}{Ne}$, and $\nucl{23}{Na}$ are also calculated by AMD. In Table~\ref{resparam},
we list the RWAs of the resonances as the ratio to the Wigner limit,  
\begin{equation}
 \theta^2_{A_1 + A_2,l} (a) = \frac{a}{3} \left|a y_l^{A_1 + A_2} (a)\right|^2,
\end{equation}
where the channel radius $a$ is chosen to connect the RWAs to the Coulomb wave function smoothly.
In principle, if the channel radius $a$ is sufficiently large so that nuclear force and channel coupling are negligible, the decay width calculated by Eq.~(6) does not depend on the choice of channel radius. However, because we performed numerical calculations within a finite spatial size, the decay widths and the reaction rates calculated from them depend on the channel radius. In order to assess this uncertainty, we also made the calculations by artificially changing the channel radius by 1 fm. We found that the reaction rates change by, at most, a factor of two. Thus, the impact of this uncertainty on astrophysical applications is relatively minor, although it cannot be ignored from a nuclear physics perspective.
The partial decay width $\Gamma_{A_1 + A_2,l}$ is calculated from the RWA,
\begin{equation}
 \Gamma_{A_1 + A_2,l} = P_\mathrm{C}^{(l)}(a) \frac{3\hbar^2}{2\mu a^2} \theta^2_{A_1 + A_2,l} (a).
\end{equation}
$P_\mathrm{C}^{(l)}$ is the Coulomb penetration factor,
\begin{equation}
  P_\mathrm{C}^{(l)} = \frac{2ka}{F_l^2(ka) + G_l^2(ka)},
\end{equation}
where $k$ is the wave number, $F_l$ and $G_l$ are the regular and irregular Coulomb functions,
respectively, and $\mu$ is the reduced mass. 

To evaluate the $S^\ast$-factor, we calculate the $\CC$ fusion cross section, assuming that each resonance is narrow and isolated. In this case, the cross section
at the center-of-mass energy $E$ is given by the Breit-Wigner formula, 
\begin{equation}
 \sigma (E) = \frac{\pi\hbar^2(2J+1)}{\mu E}
  \frac{\Gamma_0 \Gamma}{\left(E - E_\mathrm{R}\right)^2 + \Gamma^2/4},
\end{equation}
where $J$ and $E_\mathrm{R}$ denote the spin and energy of a resonance, respectively, and $\Gamma_0$
is the partial width of the entrance channel ($\CC$). The total width $\Gamma$ is estimated by a sum
over the relevant channels, 
\begin{equation}
 \Gamma = \Gamma_{a_0} + \Gamma_{a_1} + \Gamma_{a_2} + \Gamma_{p_0} + \Gamma_{p_1} + \Gamma_{p_2}.
\end{equation}
We note that this approximation may bring about additional uncertainty in the reaction rates. For instance, as seen in the results of D1M* listed in Table~1, the $2^+$ resonances at 2.10 and 2.45~MeV are close in energy and overlapping. In such cases, the shape of the cross section can deviate from that given by Eq.~(8) due to the  interference of the resonances. In principle, such resonance interference should be properly considered, but due to limitations in the framework of this study, it is neglected.

Given the cross section, the $S^\ast$-factor is defined as,  
\begin{equation}
 S^\ast(E) = E \sigma(E) / G^\ast(E),
\end{equation}
with the modified Gamow parameter $G^\ast = \exp(2\pi \eta + 0.46~\mathrm{MeV}^{-1} E)$, and the 
Sommerfeld parameter $\eta =  36/137\sqrt{\mu c^2/2 E}$ \cite{Patterson1969}.

\begin{table*}[tbp]
\caption{
 The properties of the $J^\pi=0^+$ and $2^+$ resonances obtained by using the D1M* and  SkM* density
 functionals. Only the resonances with energies ($E_R$) lower than 4 MeV are  listed. $\Gamma$ is a
 total decay  width, and $\theta^2_{\rm C}$ is a dimensionless reduced width 
 amplitude for the $\CC$ channel in percent at channel radius $a = 6$  fm (asterisked numbers are
 for $a = 7$ fm). Branching ratios larger than $10^{-2}$ are also
 listed. $M(\mathrm{IS}\lambda\uparrow)$  is the matrix element of the isoscalar monopole
 (quadrupole) transitions from the ground state of $^{24}{\rm Mg}$ to the $J^\pi=0^+$ ($2^+$)
 resonances.}  
\label{resparam}
\begin{center}
  \begin{tabular}{cccccccccccc}
  \hline
  DF & $J^\pi$ & $\er$ [MeV] & $\Gamma$ [MeV] & $\theta^2_{\rm C}$ [\%] &
		      \multicolumn{6}{c}{branching ratio} & $M(\mathrm{IS}\lambda\uparrow)$ [W.u.]   \\ 
  & & & & & $\alpha_0$ & $\alpha_1$ & $\alpha_2$ & $p_0$ & $p_1$ & $p_2$ \\
  \hline
  D1M* & $2^+$ & 0.40 & 0.14    &  0.6* & .31 & .24 &     & .19 & .25 &     & 1.37 \\
       & $0^+$ & 0.93 & 0.47    &  4.8  & .93 & .04 &     & .01 &     &     & 0.39 \\
       & $2^+$ & 1.06 & 0.10    &  1.1* & .54 & .03 &     & .27 & .14 &     & 2.34 \\
       & $2^+$ & 2.10 & 0.45    &  0.6* & .40 & .40 &     & .02 & .15 &     & 3.89 \\
       & $0^+$ & 2.33 & 0.63    &  8.8  & .58 & .38 &     &     & .03 &     & 0.74 \\
       & $2^+$ & 2.45 & 0.18    &  0.2* & .08 & .67 &     & .09 & .13 &     & 0.96\\
       & $2^+$ & 3.21 & 0.67    &  5.8  & .42 & .46 & .01 & .03 & .06 & .01 & 3.68 \\
       & $0^+$ & 3.84 & 0.73    &  8.4  & .26 & .65 & .02 & .01 & .04 &     & 1.07 \\
       & $2^+$ & 4.29 & 0.79    &  7.8  & .30 & .31 & .16 & .12 & .07 & .01 & 1.90 \\
  \hline                                                          
  SkM* & $2^+$ & 2.42 & 0.36    &  1.2* & .01 & .13 &     & .53 & .27 & .05 & 3.23 \\ 
       & $2^+$ & 3.30 & 0.38    &  1.8  & .20 & .20 &     & .20 & .36 & .01 & 1.42 \\ 
       & $0^+$ & 3.49 & 0.32    & 11.1  & .65 &     &     & .06 & .27 &     & 0.94 \\ 
       & $2^+$ & 4.27 & 2.41    &  0.5* &     &     &     & .68 & .26 & .04 & 1.83 \\ 
  \hline
\end{tabular}
\end{center}
\end{table*}

\begin{figure}[tbp]
 \begin{center}
\begin{center}
   \includegraphics[width=0.475\textwidth]{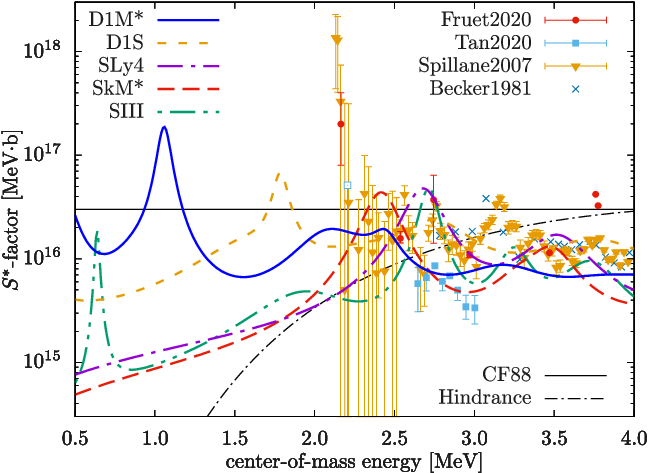}
\end{center}  
\caption{
  $\CC$ fusion S*-factors obtained by using Skyrme and Gogny density functionals. The experimental
  data~\cite{Becker1981,Spillane2007,Tan2020,Fruet2020}, the evaluations by CF88~\cite{Caughlan1988}, and the hindrance model~\cite{Jiang2007} are also
  shown.}\label{sfactor}  
 \end{center}
\end{figure}

Figure~\ref{sfactor} shows the calculated $S^\ast$-factors from all density functionals in comparison
with the observed data and extrapolations. The resonance contributions appear as peak
structures in the $S^\ast$-factor. In the $E >2.5$~MeV region, all density functionals yield the
$S^\ast$-factors with the same order of magnitude, which are roughly consistent with the observed
data, but slightly lower than the CF88. On the other hand, there is striking difference in the
low-energy region with $E \lesssim 2$~MeV. The $S^\ast$-factors obtained from the Gogny D1M* and
D1S functionals yield prominent peaks at 1 and 1.8~MeV, which enhance the $S^\ast$-factor comparable
with CF88. In contrast, the Skyrme functionals predict no significant 
peaks in this region, and their $S^\ast$-factors are closer to the hindrance model. 

This remarkable difference in the $S^\ast$-factors originates in the low-energy  $0^+$ and $2^+$
resonances, which are listed in Table~\ref{resparam} for the D1M* and SkM* functionals. For example,
a notable peak at 1~MeV in the $S^\ast$-factor obtained by the D1M* is created by the
$0^+$ and $2^+$ resonances at 0.93 and 1.06~MeV, whereas they do not exist in the SkM* results. We
remark that the contributions from the $2^+$ resonances are important. Since the peak
height of the $S^\ast$-factor created by a single resonance is proportional to 
\begin{equation}
(2J + 1) \frac{P_\mathrm{C}^{(l)}(a)}{G^\ast(\er)} \frac{\theta^2_{\rm C}(a)}{\Gamma},
\end{equation}
the $2^+$ resonance contributions are amplified by a factor of $2J + 1$ if other factors are
the same order of magnitude. This is case for the $2^+$ resonance at 1.06~MeV, and it
manifests itself as a prominent peak at 1~MeV. 

Thus, the deep sub-barrier resonances with $J^\pi=0^+$ and $2^+$ potentially have large impact on
the reaction rate, and hence, it is of utmost importance to experimentally search for them.  We have
suggested identifying the $0^+$ resonances using their strong 
isoscalar monopole transitions as a probe~\cite{Taniguchi2021}.  This approach was successfully applied in an
$\alpha$-inelastic scattering experiment~\cite{Adsley2022}, leading to the discovery of a $0^+$ resonance at
1.38~MeV. Interestingly, this resonance is located between the $0^+$ resonances obtained from
the D1M* and D1S functionals, which are located at 0.93 and 1.64~MeV, respectively. Therefore, the
reality may lie somewhere between the predictions of D1M* and D1S. Similarly, we suggest exploring 
the $2^+$ resonances, which potentially have greater impact on the reaction rate, by means of the
isoscalar quadrupole transitions. As shown in Table~\ref{resparam}, the transition strengths of the
$2^+$ resonances are greater than the Weisskopf unit, thereby demonstrating the feasibility of this
approach.  

We also comment on the partial decay widths of the $\alpha_0$ and $p_0$ channels. In the recent
particle-$\gamma$ coincidence measurements~\cite{Spillane2007,Tan2020,Fruet2020}, these widths could not be
measured. Therefore, they were estimated by assuming a linear dependence of the branching ratio on
the energy~\cite{Becker1981}. Since we do not find such clear linearity in our results, it would be
desirable to directly measure them. The $\alpha$-inelastic scattering experiments mentioned above,
in principle, can measure the partial widths.

Finally, we discuss the nuclear reaction rate $N_\mathrm{A} \braket{\sigma v}$ which is defined as a 
function of temperature $T$, 
\begin{eqnarray}
 N_\mathrm{A} \braket{\sigma v} 
 = \sqrt{\frac{8}{\pi \mu}} \frac{N_\mathrm{A} }{(k_\mathrm{B} T)^{3/2}}
  \int_0^\infty E\sigma(E)e^{-\frac{E}{k_\mathrm{B}T}} \mathrm{d}E,
\end{eqnarray}
where $N_\mathrm{A}$ and $k_\mathrm{B}$ denote the Avogadro and Boltzmann constants, respectively.
\begin{figure}[tbp]
 \begin{center}
\begin{center}
   \includegraphics[width=0.475\textwidth]{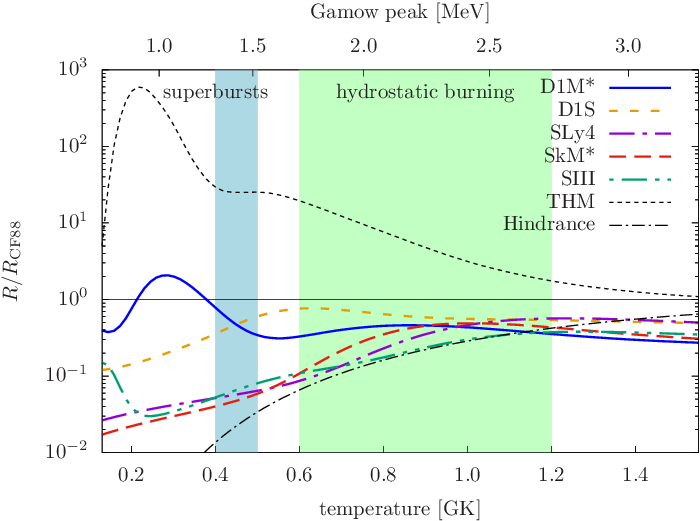}
\end{center}  
\caption{
  The $\CC$ fusion reaction rates relative to CF88 obtained by using the
  Gogny and Skyrme functionals. The upper scale shows the Gamow peak energies in MeV. The reaction
  rates obtained from the THM experiment~\cite{Tumino2018}  and the hindrance model~\cite{Jiang2007} are also
  shown.}  \label{rate}
 \end{center}
\end{figure}
Figure~\ref{rate} shows the $\CC$ fusion reaction rates relative to CF88, which exhibits the same trend
as in the $S^*$-factor. At higher temperatures, all density functionals consistently yield the
reaction rate slightly smaller than CF88. In contrast, at lower temperature $T<1$~GK, the Gogny and
Skyrme functionals predict different behavior. The former give the reaction rates slightly smaller than
CF88, whereas the latter give much hindered rates comparable with the hindrance model. We consider
that the region enclosed by the reaction rates from the Gogny and Skyrme functionals is a constraint
on the reaction rates imposed by the nuclear density functionals. The THM reaction rate is
significantly higher than this, whereas the hindrance model approximates its lower limit. 
We also remark that, very recently, Dohi et al.~\cite{Dohi2022} performed simulations of XRSB using these
reaction rates (THM, CF88 and D1S) and found differences in the results that are sufficient to
distinguish the THM rate from the others.

\section{Summary}
Using a full-microscopic nuclear model with various nuclear density functionals, we have
assessed the fusion reaction rate of $\CC$ at stellar temperatures. The reaction rates obtained from
Gogny functionals are close to the CF88
estimation. In contrast, Skyrme functionals predict lower reaction rates, similar to the
hindrance model. We interpret the intermediate region 
between the reaction rate obtained from the Gogny and Skyrme functionals as the constraints imposed
by the nuclear density functionals. 

The difference between the Gogny and Skyrme functionals owes to the
$J^\pi=0^+$ and $2^+$ resonances at deep sub-barrier energies. In the case of the Gogny functionals,
$\CC$ molecular resonances  appear approximately 1 MeV due to its coupling
with other channels, leading to a substantial enhancement of the reaction rate. On the other hand,
in the case of the Skyrme functionals, such resonances appear slightly higher energies, and  do not
contribute to the reaction rate at low temperatures.   

Hence, exploring the $J^\pi=0^+$ and $2^+$  resonances at deep sub-barrier energy is a crucial step
toward reducing the uncertainties in the reaction rate. For this purpose, we propose the
measurements of the isoscalar transitions associated with these resonances. In particular, the $2^+$
resonances, which potentially have great impact on the reaction rate, can be identified from their
enhanced isoscalar dipole transitions.

\appendix

We thank Dr.~Nishimura and Dr.~Dohi for helpful discussions.
This work was partly supported by the RCNP Collaboration Research Network program as the project number COREnet-039.
This research used computational resources of Wisteria/BDEC-01 Odyssey (the University of Tokyo), provided by the Multidisciplinary Cooperative Research Program in the Center for Computational Sciences, University of Tsukuba.
This work was financially supported by the JSPS KAKENHI Grant Nos.~22K03610 and 22H01214.


\bibliographystyle{elsarticle-num} 
\bibliography{CC_rate_plb}





\end{document}